\documentstyle[aps,here,12pt]{revtex}
\tightenlines
\def\Journal#1#2#3#4{{#1} {\bf #2}, #3 (#4)}

\def\PRL{\em Phys. Rev. Lett.}

\begin{document}
\title{
FIELD THEORETIC OPERATORS FOR MULTIFRACTAL MOMENTS
}
\author{ Ch. von FERBER$^1$ and Yu. HOLOVATCH$^2$}
\address{$^1$School of Physics and Astronomy, Tel Aviv University\\
69978 Tel Aviv, Israel\\
$^2$Institute for Condensed Matter Physics,
Ukrainian Academy of Sciences, \\ 1 Svientsitskii St.,
Lviv,  UA-290011, Ukraine}
\maketitle
\abstract{ While multifractal spectra are convex,
general field theoretic arguments show that power of
field operators $\phi^f$ yield a concave spectrum of exponents as
function of $f$. This is resolved by appropriate choice of operators
to describe multifractal moments. In a Lagrangian field theory of
two mutually interacting species of fields $\phi,\psi$, operators
$O_{f'f}=\psi^{f'}\phi^f$ with traceless symmetry give rise to
multifractal spectra of harmonic diffusion near absorbing fractals
when evaluated for zero component fields.
}
%============================================================================
%                          CHAPTER  I
%============================================================================
\section{Introduction}\label{I}
The concept of multifractality developed in the last decade has proven
to be a powerful tool for analyzing systems with complex statistics
which otherwise appear to be intractable.\cite{Hentschel83,Halsey86}
It has found direct applications in a
wide range of fields including turbulence, chaotic attractors,
Laplacian growth phenomena
etc.\cite{Halsey86,Halsey86a,Witten86,Meakin88,Marsili91,Deutsch94,Eyink94}
Here we generalize the idea of Cates  and Witten \cite{Cates87,Cates87a}
by deriving the multifractal (MF) spectrum in
the frames of a field theoretical (FT) formalism and make use of
renormalization group (RG) methods.  We  relate the MF spectrum  to
the spectrum of scaling dimensions of a family of composite operators
of Lagrangian $\phi^4$ field theory.
This gives an example of power of field
operators whose scaling dimensions show the appropriate
convexity for a MF spectrum,\cite{Duplantier91}
while there is no need to include field gradients for this property.
We calculate the MF spectrum to third order of
perturbation theory using two complementary approaches: zero mass
renormalization with successive $\varepsilon$-expansion (see
e.g.\cite{Brezin76}) and massive renormalization group approach at
fixed dimension,\cite{Parisi80} reproducing previous results
obtained in lower order of perturbation theory for special
cases.\cite{Cates87,Cates87a}
The resulting series are asymptotic. We take this into account and
obtain numerical values only by careful resummation.

We address a special case of a  growth process controlled by a
Laplacian field.
The latter may describe a variety of phenomena depending on the
interpretation of the field.
For diffusion limited aggregation this
field is given by the concentration of diffusing particles, in
solidification processes it is given by the temperature field, in
dielectric breakdown it is the electric potential, in viscous fingers
formation it is the pressure.\cite{Meakin88,Family88} In all
mentioned phenomena the resulting structure appears to be of fractal
nature and is characterized by appropriate fractal
dimensions.\cite{Hentschel83} Its growth and spatial
correlations lead to  (non-trivial) spectra of
multifractal dimensions.\cite{Halsey86} In general, the boundary
conditions determining the field will be given on the surface of the
growing aggregate itself. It is this dynamic coupling that produces the
rich structure of the phenomena and seems to make the general
dynamical problem intractable.

Here we study the simpler case when the fractal has been already formed and
look for the distribution of the Laplacian field and its higher
moments near the surface of the fractal.\cite{Cates87,Cates87a}
We will follow the diffusion
picture, considering the aggregate as an absorbing fractal,
``the absorber''. The field $\rho(\vec{r})$ gives the concentration of
diffusing particles and vanishes on the surface of the absorber.
More specifically we consider the Laplacian field $\rho(\vec{r})$ in
the vicinity of an absorbing (fractal) path, `a polymer', or a junction
of absorbing paths, `a core of a polymer star'. In general we assume
the ensemble of absorbers to be characterized by either random walk
(RW) or self-avoiding walk (SAW) statistics.
Multifractal scaling is found for the $f$-moments
$\langle \rho^f(\vec{r})\rangle$
of the field with respect to these ensembles.

This formulation of the problem allows us to use the polymer picture and
theory developed for polymer networks and
stars \cite{Duplantier86,Duplantier89,Schaefer92} and
extended for copolymer stars.\cite{FerHol96e,FerHol96c} The theory
is mapped to a Lagrangian $\phi^4$ field theory with
several couplings \cite{Schaefer85,Schaefer90,Schaefer91}
and higher order composite operators
to describe star vertices.

In section 2 we
present the path integral solution of the Laplace equation and relate
it to polymer theory. Field theoretical representation and renormalization are
discussed in section \ref{III} together with the renormalization group flow
and expressions for the exponents. In section \ref{IV} we
define the multifractal spectrum and give its series expansion
 in both renormalization group approaches adopted here.
Section \ref{V} is devoted to resummation of these asymptotic series and
numerical results followed by some
conclusions and an outlook in section \ref{VI}.

%============================================================================
%                          CHAPTER  II
%============================================================================
\section{Path Integral Solution of the Laplace Equation and
Polymer Absorber Model
}\label{II}
In this section we describe the diffusion of
particles in the vicinity of a polymer absorber by a ``polymer''
formalism. Let us formulate
the problem first in terms of diffusion of particles in time. The
probability of finding a randomly walking particle at point $\vec{r}_1$
at time $t$ which started at point $\vec{r}_0$ at time $t=0$ is given
described by the following normalized path integral:
\begin{equation}\label{2.1}
G^{0}(\vec r_0,\vec r_1,t) = \langle \delta(\vec{r}^{(1)}(0)-\vec{r}_0)
\delta(\vec{r}^{(1)}(t)-\vec{r}_1)\rangle_{{\cal{H}}_0(t)}.
\end{equation}
Angle brackets in (\ref{2.1}) denote the normalized integral
\begin{equation}\label{2.2}
\langle \dots \rangle_{{\cal{H}}_0(t)} =
\frac{\int (\dots) \exp (-  {\cal{H}}_0(t)){\rm d} \{ r^{(1)} \}}
{ \int \exp (-  {\cal{H}}_0(t))
{\rm d} \{ r^{(1)} \}},
\end{equation}
which is performed with the Hamiltonian:
\begin{equation}\label{2.3}
{\cal{H}}_0(t)=\int_0^t\Big(\frac{{\rm d}\vec{r}^{(1)}(\tau)}{{\rm d}
\tau}\Big )^2 .
\end{equation}
Integration in (\ref{2.1}) is performed over all paths $\vec{r}^{(1)}(\tau)$
with $0 \leq \tau \leq t$. Note that we have absorbed the diffusion
constant into re-definition of time.  $G^{0}(\vec r_0,\vec r_1,t)$ obeys the
following differential equation:
\begin{equation}\label{2.4}
\Big( \Delta_{r_1} + \frac{d}{2t} - \frac{\partial}{\partial t} \Big )
G^{0}(r_0,r_1,t)  = 0
\end{equation}
with $d$ the dimension of space. For finite $d$ the random walker will visit
any site after some finite time and we may assume a steady state limit
for $G^{0}(r_0,r_1,t)$ for $t=\infty$.
In this case $G^{0}(\vec r_0,\vec r_1,t)$ will become independent of
$\vec{r}_0$ and its limit and defines a field:
\begin{equation}\label{2.5}
\rho(\vec r_1)=\lim_{t \rightarrow \infty }\frac {1}{V}
\int {\rm d} \vec{r}_0 G^{0}(r_0,r_1,t),
\end{equation}
here $V$ is the system volume. The field $\rho(\vec r)$ obeys the
Laplace equation:
\begin{equation}\label{2.6}
\Delta \rho (\vec r)= 0.
\end{equation}

We introduce boundary conditions in such a way that the field
$\rho (\vec r)$ equals to some constant $\rho_{\infty}$ at $r=\infty$ and
vanishes on the absorber. The absorber itself we describe
by a path $\vec r^{(2)}(s)$,
$0\leq s\leq S_2 $.

Let us explain the solution of the Laplace equation (\ref{2.5})
in the presence of an absorbing path
$\vec{r}^{(2)}(s)$. The boundary conditions are implemented by an
avoidance interaction $u_{12}$ punishing any coincidence of the path
$r^{(1)} $ of the RW and the path $ r^{(2)}$ of the absorber.
The correlation function of a random walk in the presence of an
absorbing path $\vec{r}^{(2)}(s)$
may then be written as
\begin{eqnarray}
G(\vec{r}_0,\vec{r}_1,S_1)=\langle
\delta (\vec{r}^{(1)}(0) - \vec{r}_0)
\delta (\vec{r}^{(1)}(S_1) - \vec{r}_1)
\nonumber \\
\exp \Big \{
- u_{12} \int_0^{S_1}{\rm d}s_1 \int_0^{S_2}{\rm d}s_2
 \delta (\vec{r}^{(1)}(s_1) - \vec{r}^{(2)}(s_2)
\Big \}
\rangle_{{\cal{H}}_0(S_1)} ,
\label{2.10}
\end{eqnarray}
where we have adopted the notation $t=S_1$.

We are interested in ensemble moments $\langle\rho^{f_2}(\vec r_1)\rangle$
of the field in the vicinity of the absorber, assuming an ensemble
of RW or SAW absorbers.
For the RW ensemble the average is performed with respect to the
Hamiltonian ${\cal H}_0(S_2)$, for the SAW ensemble an additional
interaction has to be included.

Here we choose a more general formulation, which allows us to describe
the moments of the field in the vicinity of the core of an absorbing
polymer star, or near the junction of $f_1$ absorbing paths.

The calculation of the $f_2$ moment of (\ref{2.10}) near the junction
of $f_1$ absorbing paths will include the average over $f_2$ random
walks ending at $r_1$ and the ensemble average over the configurations
of the $f_1$ absorbing paths with junction at $r_1$.

 We are thus lead to consider the partition function of
a star of walks  which are in part mutually-avoiding. We
will give this partition function here for the more general case of
$f$ walks of which $f_2$ random walks describe the field and $f_1$
walks correspond to the absorber. This situation describes the
$f_2$th  moment of the flux of the field to the core of an absorbing
star of $f_1$ walks. We allow for additional avoidance interactions
among these absorbing paths:
\begin{eqnarray}\label{2.11}
{\cal Z}_{*f}&=& \frac{1}{{\cal Z}^0_{*f}}
\langle \prod_{i=1}^{f}
\delta (\vec{r}^{(i)}_i(0) - \vec{r}_0)
\exp \Big \{
-\frac{1}{6} \sum_{a,b=1}^{f} u_{ab}
\int_0^{S_a}{\rm d}s_a \int_0^{S_b}
{\rm d}s_b
\nonumber \\
&&
\delta (\vec{r}^{(a)}(s_a) - \vec{r}^{(b)}(s_b)
\Big \}
\rangle_{{\cal{H}}_0^{*f}\{S_a\}} ,
\end{eqnarray}
here
$$
{\cal{H}}_0^{*f}\{S_a\}=\sum_{i=1}^{f}{\cal{H}}_0(S_i)
$$
where ${\cal{H}}_0(S_i)$ is given by (\ref{2.2}).
$Z^0_{*f}$ stands for the partition function of star with zero
interactions $u_{ab}=0$. The matrix $u_{ab}$ is given in the following
form:
\begin{equation}
u_{ab} = \left \{ \begin{array}{ll}
u_{11}^0 & \mbox{if $a,b \leq f_1$}
\\
u_{12}^0 & \mbox{if $a \leq f_1 < b \leq f_1+f_2$}
\\
   & \mbox{or $b \leq f_1 < a \leq f_1+f_2$}
\\
0  & \mbox{else}
\end{array} \right.
\end{equation}
This corresponds to the partition function of co-polymer stars
consisting of two species of chains \cite{FerHol96e,FerHol96c}
with $f_1$ chains of one species and $f_2$ chains of the
other. $u_{11}^0$ is the interaction between absorbing paths.

%============================================================================
%                          CHAPTER  III
%============================================================================
\section{Field Theory and Renormalization
}\label{III}
As  is well known, the polymer model may be mapped to the limit of
$m=0$ of  $O(m)$-symmetrical Lagrangian field theory.\cite{deGennes72}
To describe polymers and interacting random walks
at the same time we adopt the formalism developed for multicomponent
polymer solutions.\cite{Schaefer91} Its field theory  is described
by the following Lagrangian:
\begin{eqnarray}\label{3.1}
\lefteqn{ {\cal L}\{\phi_b,\mu_b\} = \frac{1}{2}
\sum_{a=1}^{f}\int{\rm d}^d r \left(\mu_a\phi_a^2 + (\nabla \phi_a(r)
)^2 \right)}&&\nonumber\\ &&+ \frac{1}{4!} \sum_{a,a^{'}=1}^{f}
u_{a,a'} \int {\rm d}^d r \phi_a^2(r)\phi_{a'}^2(r) .
\end{eqnarray}
in general $m$-component theory
\begin{equation}\label{3.2} \phi_a^2 = \sum_{\alpha = 1}^{m}(
  \phi_a^{\alpha} )^2 .  \end{equation}
$\mu_a$ is a chemical potential conjugated to the Gaussian surfaces
$S_a$ in (\ref{2.10}). Correlation functions in this theory are
defined by averaging with the weight given by (\ref{3.1}):
\begin{equation} \label{3.3}
\langle (\dots) \rangle |_{{\cal L}}
=   \int{\cal D}[\phi_a(r)] (\dots)
   \exp[-{\cal L}\{\phi_b,\mu_b\}] \, |_{m=0}.
\end{equation}
here functional integration $\int{\cal D}[\phi_a(r)]$ is defined in
such a way that normalization is already included:
$\langle 1 \rangle |_{{\cal L}} = 1 $ if all $u_{a,a^{\prime}}=0$.
The limit $m=0$ in (\ref{3.3}) can be understood as a certain rule to
calculate the diagrams appearing in the perturbation theory expansions
and can be easily checked diagrammatically.

The partition function ${\cal Z}_{*f}$ defined  in (\ref{2.10}) is mapped to the
field theoretical correlation function $\tilde{\cal Z}_{*f}$ via a Laplace
transform in the Gaussian surfaces $S_a$ to conjugate chemical potentials
(``mass variables'') $\mu_a$:
\begin{equation} \label{3.4}
\tilde{\cal Z}_{*f}\{\mu_a\} = \int _0^{\infty}
\prod_b {\rm d}S_b e^{-\mu_b S_b} {\cal Z}_{*f}\{ S_a \} ,
\end{equation}
and
\begin{equation} \label{3.5}
\tilde{\cal Z}_{*f}\{\mu_a\} = \langle
\int {\rm d} r_a \prod_{a=1}^{f} \phi_a(r_0) \phi_a(r_a)
\rangle |_{{\cal L}}
\end{equation}

Our interest is in the scaling properties of these
functions.
Note that by (\ref{3.5}) these are governed by the spectrum of scaling
dimensions of the composite operators $\prod_{a=1}^{f} \phi_a$.
To extract them we use renormalization group
methods.\cite{Bogoliubov59,Zinn89} Here we use the results of
our previous approaches  to the problem of co-polymer stars: massless
renormalization group scheme with successive $\varepsilon$-expansion
(see e.g.\cite{Brezin76}) and massive renormalization group approach
at fixed dimension \cite{Parisi80} compiled in a
pseudo-$\varepsilon$ expansion.\cite{pseps}
 On the basis of
correlation functions it is standard to define vertex functions
$\Gamma^{4}_{u_{ab}}$ corresponding to the couplings $u_{ab}$ as well
as vertex functions $\Gamma^{*f}_{\Pi \phi_a}$ with insertion of
composite operators $\prod_a \phi_a$. Explicit
expressions may be found in \cite{FerHol96e,FerHol96c}. We define
renormalization and introduce renormalized couplings $g_{ab}$ by:

\begin{equation} \label{3.6}
u_{ab} = \mu^{\varepsilon} Z_{\phi_a}Z_{\phi_b}Z_{ab} g_{ab} .
\end{equation}
The renormalizing $Z$-factors are power series in $g_{ab}$ according
to the following conditions:
\begin{eqnarray}
Z_{\phi_a}(g_{aa}) \frac{\partial}{\partial k^2}
\Gamma_{aa}^{(2)}(u_{aa}(g_{aa})) = 1                   \label{3.7}\\
Z_{ab}(g_{ab})
\Gamma_{aabb}^{(4)}(u_{ab}(g_{ab})) = \mu^\varepsilon g_{ab}
\label{3.8}
\end{eqnarray}
$\mu$ is a scale parameter (equal to the mass at which the massive
scheme is evaluated and giving the scale of external momenta in the
massless scheme).

In order to renormalize the star vertex functions we introduce
renormalization factors  $Z^{*f}_{\Pi \phi_a}$ by:
\begin{equation} \label{3.9}
(\prod_{a=1}^{f} Z_{\phi_a}^{1/2}) Z^{*f}_{\Pi \phi_a}
\Gamma^{*f}_{\Pi \phi_a}(u_{ab}(g_{ab})) = \mu^{\delta_{\Pi \phi_a}},
\end{equation}
where $\delta_{\Pi \phi_a}$ is the engineering dimension of the
composite operator
\begin{equation} \label{3.10}
\delta_{\Pi \phi_a}\ = \ f(\frac {\varepsilon}{2} -1)
+ \ 4 - \ \varepsilon
\end{equation}
The dependence of the renormalized couplings $g_{ab}$ and of
renormalizing $Z$-factors on the scale parameter $\mu$ is expressed by
the following relations:
\begin{eqnarray}
\mu \frac{\rm d}{{\rm d}\mu} g_{ab} &=& \beta_{ab}(g_{a'b'}) ,
\label{3.11} \\
\mu \frac{\rm d}{{\rm d}\mu} \ln Z^{*f}_{\Pi \phi_a}(g_{ab}) &=&
\eta_{\Pi \phi_a}(g_{ab}) .
\label{3.12}
\end{eqnarray}
We are going to look on the situation of having two sets of walks of
different species. In this case only three different couplings remain.
We will refer to them as $g_{11}$, $g_{22}$, $g_{12}=g_{21}$.
The corresponding functions $\beta_{11}$, $\beta_{22}$, $\beta_{12}$ define
a flow in the space of couplings. This renormalization group flow was
discussed in \cite{Schaefer90,Schaefer91}. Its fixed points are
determined by a set of equations:
\begin{equation} \label{3.13}
\beta_{ab}(g_{ab}^*)=0, \hspace{2em} a,b=1,2.
\end{equation}

In the space of the three couplings one finds \cite{Schaefer91} 8
fixed points corresponding to absence or presence of inter- and intra-
species interaction. They are given in the table \ref{tab1} where
$g^*$ corresponds to the fixed point of the theory containing only 1
species, whereas $g_G^*$ corresponds to the case of having only
inter-species interactions, $g_U^*$ describes a set of random walks
interacting with another set of self-avoiding walks.
The phenomenon we address in this article corresponds to the case of
non-vanishing interaction between the two species of walks,
while one set has no self-interaction.
 Thus we consider the
two fixed points labeled {\it G}
and {\it U}. The first corresponds to a set of random walks
interacting with random walks of another species and thus describes
absorption on random walk absorbers, the second corresponds to a
set of random walks interacting with another set of self-avoiding
walks
and thus describes absorption on SAW (polymer) absorbers.

Having $f_1$ walks of the first species and $f_2$ walks of second
species we define the following exponents in the fixed points
{\it G,U}:
\begin{eqnarray}
\label{3.14}
\eta^{G}_{f_1f_2}&=& \eta_{\Pi \phi_a}(g_{11}=g_{22}=0, g_{12}=
g^*_{G}),
\\
\label{3.15}
\eta^{U}_{f_1f_2}&=& \eta_{\Pi \phi_a}(g_{11}=g^*, g_{22}=0, g_{12}=
g^*_{U}),
\end{eqnarray}
which govern the scaling properties of the partition sum (\ref{2.11}).

The scaling may be formulated in terms of the size $R$ of the walks:
We have to normalize the
partition function by the number of configurations of the absorber
given by  ${\cal Z}_{*f_10}$.
For large $R$ the resulting quantity scales like
\begin{equation}
\label{3.16}
{\cal Z}_{*f_1f_2}/{\cal Z}_{*f_10} \sim R^{-\lambda_{f_1f_2}}.
\mbox{\hspace{2em} for } R=S^{\nu} \rightarrow \infty
\end{equation}
Here $\lambda_{f_1f_2}=\eta_{f_1f_2}-\eta_{f_10}$, $\nu$ is
the correlation length critical exponent of the walks:
$\nu=1/2$ for random walks and $\nu \simeq 0.588$ for self-avoiding
walks at $d=3$. For the fixed point {\it G} we have $\eta_{f_10}^G=0$ and
$\nu=1/2$ for all walks.

The exponent $\lambda_{2,n}$ corresponds to the $n$th moment of the
flux onto the center segment of an
absorbing linear chain. Considering the absorber to be
either a random walk or a self-avoiding walk let us define the exponents:
\begin{eqnarray}
\label{3.17}
\lambda_{RW}(n) \equiv \lambda_{2,n}^G=-\eta^{G}_{2n},
\\
\label{3.18}
\lambda_{SAW}(n) \equiv \lambda_{2,n}^U=-\eta^{U}_{2n}+\eta_{20}
\end{eqnarray}

Previously \cite{FerHol96e,FerHol96c} we obtained
the expressions for the exponents $\eta^{G}_{f_1f_2}$,
$\eta^{U}_{f_1f_2}$ in terms of $\varepsilon$-expansion and
pseudo-$\varepsilon$ expansion series in massless
and massive renormalization group schemes. Whereas the first
corresponds to collecting perturbation theory terms of the same powers
of $\varepsilon=4-d$, in the pseudo-$\varepsilon$ expansion
series \cite{pseps} at each power of the
pseudo-$\varepsilon$ parameter ($\tau$)
one collects contributions from the dimension-dependent loop integrals
of the same order.
In the final results $\tau=1$.
Based on the expressions for the exponents $\eta^{G}_{f_1f_2}$,
$\eta^{U}_{f_1f_2}$ \cite{FerHol96e,FerHol96c} we find:
\begin{eqnarray}
\lambda_{RW} (\varepsilon) &=&
\varepsilon\,n-{\frac {n\left (n-1\right ){\varepsilon}^{2}}{4}}+{\frac
{n\left (n-1\right )\left (-1+n+3\,\zeta (3)\right ){\varepsilon}^{3}
}{8}}
\label{3.23}
\end{eqnarray}
\begin{eqnarray}
\lambda_{SAW} (\varepsilon) &=&
{\frac {3\,\varepsilon\,n}{4}} +
\left ({\frac {7\,n}{
128}}-{\frac {9\,{n}^{2}}{64}}\right ){\varepsilon}^{2}+
 \nonumber \\ &&
\left (-{\frac {149\,n}{2048}}-{\frac {21\,{n}^{2}}{1024}}+{\frac
{27\,{n}^{3}}{512}}-{\frac {69\,n\zeta (3)}{512}}+{\frac {135\,{n}
^{2}\zeta (3)}{512}}\right ){\varepsilon}^{3}
\label{3.24}
\end{eqnarray}
\begin{eqnarray}
\lambda_{RW} (\tau) &=&
\tau\,\varepsilon\,n
+\left ({\frac {\varepsilon\,{n}^{2}}{2}}-\varepsilon\,{n}^{2}{\it i_1}-
{\frac {\varepsilon\,n}{2}}+\varepsilon\,n{\it i_1}\right ){\tau}^{2} +
\lambda_{RW}^{3 loop} {\tau}^3.
\label{3.25}
\end{eqnarray}
\begin{eqnarray}
\lambda_{SAW}(\tau)&=&
{\frac {3\,\tau\,\varepsilon\,n}{4}}
+\left ({\frac {\varepsilon\,n{\it i_1}}{4}}+
{\frac {9\,\varepsilon\,{n}^{2}}{32}}-
{\frac {9\,\varepsilon\,{n}^{2}{\it i_1}}{16}}+
{\frac {\varepsilon\,n{\it i_2}}{16}}-
{\frac {\varepsilon\,n}{8}}\right ){\tau}^{2}+
\nonumber \\
&&\lambda_{SAW}^{3 loop} {\tau}^3.
\label{3.26}
\end{eqnarray}

Here $\zeta(3) \simeq 1.202$ is the Riemann zeta function,
$i_j$ are the loop  integrals dependent on the space dimension $d$:
at $d=3$ $i_1=2/3$, $i_2=-2/27$.
The expressions for the three-loop terms $\lambda_{RW}^{3 loop}$,
$\lambda_{SAW}^{3 loop}$ in (\ref{3.25}),(\ref{3.26})
are given elsewhere.\cite{FerHol96d}
%============================================================================
%                          CHAPTER  IV
%============================================================================
\section{Multifractal Spectrum}\label{IV}
A widely used description for the MF spectrum is obtained from a
Legendre transform, the spectral function $f(\alpha)$, of the analytically
continued spectrum $\lambda(n)-n\lambda(1)$ by
\begin{equation}
f(\alpha)= (\lambda(n)-n\lambda(1)) + n \alpha
\mbox{\hspace{2em} with \hspace{2em}}
\alpha=-\frac{\rm d}{{\rm d}n}(\lambda(n)-n\lambda(1))
\end{equation}
For standard moments of a MF measure not including an ensemble average
$f(\alpha)$ is called the spectral function. Here this notion is kept.
The spectral function is widely used to characterize the multifractal
nature of many processes.\cite{Halsey86}
In the standard approach the function
$f(\alpha)$ defined for a multifractal measure on a set $X$ gives for
every $\alpha$ the fractal dimension of the subset of $X$ for which
the measure at scale $\ell$ is characterized by $\ell^\alpha$ with the
H\"older exponent $\alpha$, in the limit $\ell\to 0$.
Due to this interpretation as a property of a (multifractal) measure,
strict convexity conditions hold. The standard
$f(\alpha)$ has the shape of a cap. While simple power of field
operators $\phi^f$ will not generate such a
spectrum,\cite{Duplantier91} the operators constructed here accord to
this condition.

Using the perturbation expansions for the $\lambda$ exponents given
to third loop order both in $\varepsilon$ and $\tau$ expansion in
massless and massive renormalization (\ref{3.23}) - (\ref{3.26})
and the relations for $\lambda(n)$
and the spectral function some algebra leads to the corresponding
expansions for $\alpha_n$ and $f(\alpha_n)$:
\begin{eqnarray}
\alpha_{RW} (\varepsilon) &=&
2 + \left (-{\frac {n}{2}}+1/4\right ){\varepsilon}^{2} +
\nonumber\\ & &
\left (-{\frac {n}{2}}-{\frac {3\,\zeta (3)}{8}}+{\frac {3\,{n}^{2
}}{8}}
+{\frac {3\,n\zeta (3)}{4}}+1/8\right ){\varepsilon}^{3}.
\label{alpharwe}
\end{eqnarray}
\begin{eqnarray}
f_{RW} (\varepsilon) &=&
2  - {\frac {{\varepsilon}^{2}{n}^{2}}{4}} +
\left ({\frac {{n}^{3}}{4}}+{\frac {3\,{n}^{2}\zeta (3)}{8}}-
{\frac {{n}^{2}}{4}}\right ){\varepsilon}^{3}
\label{frwe}
\end{eqnarray}
\begin{eqnarray}
\alpha_{RW} (\tau) &=&
2+ \left (-{\frac {\varepsilon}{2}}+\varepsilon\,{\it i_1}+
\varepsilon\,n-2\,\varepsilon\,n{\it i_1}\right ){\tau}^{2}+
\alpha_{RW}^{3loop}\tau^{3}
\label{alpharwt}
\end{eqnarray}
\begin{eqnarray}
f_{RW} (\tau) &=&
2+\left ({\frac {\varepsilon\,{n}^{2}}{2}}-
\varepsilon\,{n}^{2}{\it i_1}\right ){\tau}^{2} +
f_{RW}^{3loop}\tau^{3}
\label{frwt}
\end{eqnarray}
\begin{eqnarray}
\alpha_{SAW} (\varepsilon) &=&
2 - 1/4 \varepsilon + \Big (
{\frac {7}{128}}-{\frac {9\,n}{32}} \Big ) \varepsilon^2
\nonumber \\ &&
\Big (
-{\frac {149}{2048}}-{\frac {69\,\zeta (3)}{512}}-{\frac {21\,n}{
512}}+{\frac {81\,{n}^{2}}{512}}+{\frac {135\,n\zeta (3)}{256}}
\Big ) \varepsilon^3
\label{alphasawe}
\end{eqnarray}
\begin{eqnarray}
f_{SAW} (\varepsilon) &=&
2 - 1/4 \varepsilon +
\Big (
-{\frac {9\,{n}^{2}}{64}}-{\frac {11}{128}}
\Big ) \varepsilon^2
\nonumber \\ &&
\Big (
{\frac {27\,{n}^{3}}{256}}+
{\frac {135\,{n}^{2}\zeta (3)}{512}}-
{\frac {21\,{n}^{2}}{1024}}-
{\frac {83}{2048}}+
{\frac {33\,\zeta (3)}{256}}
\Big ) \varepsilon^3
\label{fsawe}
\end{eqnarray}
\begin{eqnarray}
\alpha_{SAW} (\tau) &=&
2 -{\frac {\varepsilon}{4}} \tau
+ \Big (
-{\frac {\varepsilon}{8}}-
{\frac {9\,\varepsilon\,n{\it i_1}}{8}}+
{\frac {9\,n\varepsilon}{16}}+
{\frac {\varepsilon\,{\it i_1}}{4}}+
{\frac{\varepsilon\,{\it i_2}}{16}}
\Big ) \tau^2 +
\nonumber \\ &&
\alpha_{SAW}^{3loop}\tau^{3}
\label{alphasawt}
\end{eqnarray}
\begin{eqnarray}
f_{SAW} (\tau) &=&
2
-{\frac {\varepsilon}{4}} \tau +
\Big (
-{\frac {9\,\varepsilon\,{n}^{2}{\it i_1}}{16}}+
{\frac {9\,\varepsilon\,{n}^{2}}{32}}+
{\frac {5\,\varepsilon}{32}}+
{\frac {\varepsilon\,{\it i_2}}{16}}-
{\frac {5\,\varepsilon\,{\it i_1}}{16}}
\Big ) \tau^2 +
\nonumber \\ &&
f_{SAW}^{3loop}\tau^{3}
\label{fsawt}
\end{eqnarray}

Here $\zeta(3) \simeq 1.202$ is the Riemann zeta function,
$i_j$ are the loop  integrals dependent on the space dimension $d$:
at $d=3$ $i_1=2/3$, $i_2=-2/27$.

Again $\zeta(3)$ is the Riemann zeta function, $i_1$ and $i_2$ are the
two-loop integrals and the explicit form of the three-loop
contributions in (\ref{alpharwt}), (\ref{frwt}), (\ref{alphasawt}),
(\ref{fsawt}) is given elsewhere.\cite{FerHol96d}

%============================================================================
%                          CHAPTER  V
%============================================================================
\section{Resummation
}\label{V}

As is well known, the series of type (\ref{3.23})-(\ref{3.26}),
(\ref{alpharwe}) - (\ref{fsawt}), as they occur in field theory appear
to be of asymptotic nature with zero radius of convergence. However,
knowledge of the asymptotic behavior of the series as
derived from the renormalization group theory allows us to evaluate these
asymptotic series (see e.g.\cite{LeGuillou80}). To this end several
procedures are available depending on the additional information known
for the series to be resummed. We assume our series to be of
$\varepsilon$-expansion character
together with additional information  for the case of
the Lagrangian \cite{Schaefer91} (\ref{3.1}) we consider here. So we expect the
following behavior of the $k$th order perturbation theory term $A_k$
for any of given above quantities:  \begin{equation} \label{5.1} A_k
\,\sim \,k!\,k^b\,(-a)^k \end{equation} the constant $a$ for the
$\varepsilon$-expansion of Lagrangian $\phi^4$ field theory with one
coupling was derived in  \cite{Lipatov77,Brezin77}:  $a=3/8$. For the
unsymmetric fixed point, where two different couplings are present we
use the value $a=27/64$.\cite{Schaefer91} We assume as well that the
same properties hold also for the pseudo-$\varepsilon$ expansion in
terms of $\tau$. With the above information in hand one can make use
of the Borel summation technique improved by the conformal mapping
procedure which up to now served a powerful tool in the field theory
(see \cite{LeGuillou80} for example).

\begin{figure} [H]
\begin{center}
(Fig.1a)\input ferhol1.pic\\[5mm]
(Fig.1b)\input ferhol2.pic
\end{center}
\caption{ \label{fig2}
Spectral function $f(\alpha)$ for absorption on (a) $RW$ and (b) $SAW$.
Solid curves: 1 - [2/1] Pad\'e approximant for $\varepsilon^3$ results,
2- [2/1] Pad\'e approximant for pseudo-$\varepsilon^3$ results;
dashed curves: 3 - $\varepsilon^2$ results without resummation,
4 - pseudo-$\varepsilon^2$ results without resummation;
stars - resummed $\varepsilon^3$ results;
boxes - resummed pseudo-$\varepsilon^3$ results.
}
\end{figure}
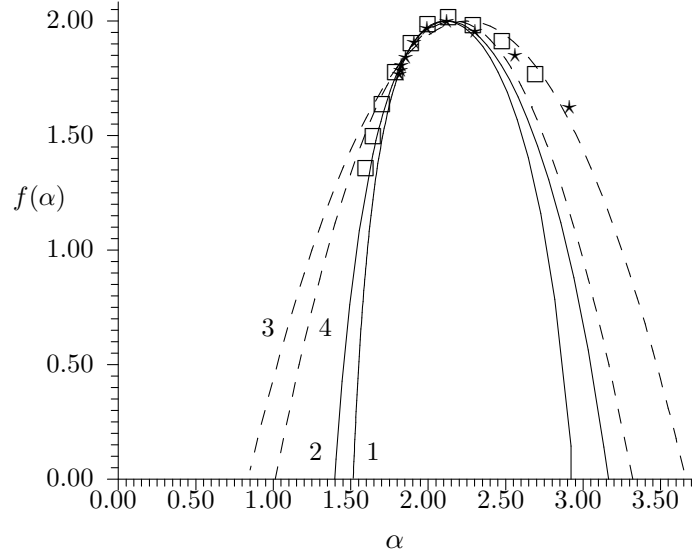
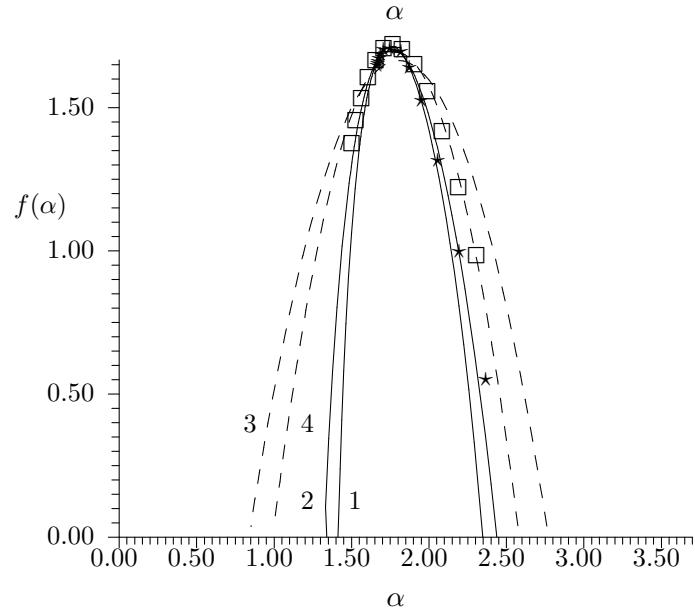

Table \ref{tab2} contains the results for the exponents $\lambda_{RW}(n)$
and $\lambda_{SAW}(n)$ obtained in $\varepsilon$ and in
pseudo-$\varepsilon$ expansion techniques from the corresponding
values of exponents $\eta_{f_1,f_2}$  \cite{FerHol96e}
with the application of the
resummation procedure as described above.
We have calculated the spectral function as a Legendre transform from the
series of exponents received from the scaling of the moments of measure
defined by diffusion. These moments were calculated  as averages over
all configurations of the absorber instead of performing a site average.
Thus the interpretation of $f(\alpha)$ itself does not
directly correspond to the picture developed above for the standard MF.
All the same deriving $f(\alpha)$ in the same way from the spectrum of
scaling exponents of moments of a measure  the above discussed properties
of a $MF$ spectral function hold also for $f(\alpha)$ in this case.
In particular $\max f(\alpha) = \tau_0$ gives the fractal dimension
of the absorber and $f$ is convex $f''(\alpha)< 0$.
Our numerical results for the spectral function
are presented in Figs. \ref{fig2}a,\ref{fig2}b.
They were obtained from the series for $\alpha_n$ and $f(\alpha_n)$ as
functions of $n$. We show the results of the
resummation procedure described above applied to the
series in both RG approaches.  For comparison we also show the curve
for direct summation of the $\varepsilon$ and $\tau$ series to the
2nd order. In addition we have performed an analytical continuation of
our series in form of [2/1] Pad\'e approximants for the
$\varepsilon^3$ and $\tau^3$ series.  It is obvious that direct
summation of $\varepsilon^3$ and $\tau^3$ series
fails to converge and gives comparable values for
$\alpha_n,f(\alpha_n)$ only for small values of $n$, i.e. near the
maximum of $f(\alpha)$ at $n=0$.  The symmetry of the Pad\'e
approximant holds only in the region shown and may be an artifact of
the method. On the left wing, where it coincides with the resummed
results the Pad\'e approximant gives a continuation which is
compatible with the estimation for the minimal $\alpha$ value
$\alpha_{\min}= d-2$. The Pad\'e result is
$\alpha_{\min}(\varepsilon)=1.333$, $\alpha_{\min}(\tau)=1.017$
for the RW absorber and
$\alpha_{\min}(\varepsilon)=1.250$, $\alpha_{\min}(\tau)=1.013$
for the SAW absorber, which is calculated here only from 3rd order
perturbation theory.

Note that though the
results obtained for $\alpha_n$ and $f(\alpha_n)$ for a specific value of $n$
differ in both approaches, the same curve $f(\alpha)$ is described with
better coincidence for the left wing of the curves, corresponding to
positive $n$.
The resummation techniques we apply have proven to be a powerful tool
already in the field theoretical approach to critical phenomena and have
lead to high precision values for critical exponents.
We hope that the application of these methods to calculations of MF
phenomena allows to improve the reliability and comparability of the
results.

We currently work on the generalization of our present approach to find
the series of spectral functions obtained for absorption at the core
of a polymer with any number of arms.

As can be extrapolated from the Pad\'e approximant and as was shown also
on the basis of high order approximations,\cite{Cates87} $f(\alpha)$
as it is defined here, will become negative near $\alpha_{\min}$ and
$\alpha_{\max}$. For this reason the identification of $f(\alpha)$ as the
fractal dimension of some identifiable subset is not possible here.
Also the extrapolation of the resummed data seems to indicate such a
behavior. Note, however, that the perturbative approach, even in combination
with resummation and analytical continuation is still not capable to give
reliable results for high values of the expansion parameters. In
particular this method is only good near the maximum of $f(\alpha)$.

The possible negative values of the spectral function were discussed
already in \cite{Cates87} and a physical interpretation of $f(\alpha)$
was given as a histogram of the measure $\mu$ plotted in logarithmic
variables. In this interpretation negative $f(\alpha)$ indicate that
the number of sites with a certain logarithmic measure
$\alpha\sim\ln\mu$ decreases as the size of the absorber $R$
increases. Thus for large $R$ this number can only be defined by an
ensemble average, as performed here.

%============================================================================
%                          CHAPTER  VI
%============================================================================
\section{Conclusions
}\label{VI}

We have studied the characteristics of harmonic diffusion in the
presence of a fractal absorber.
We related the description of diffusing particles near absorbing paths
to interacting walks. Following the model proposed by Cates and
Witten \cite{Cates87} we used the polymer formalism to describe both the
absorber and the random walks of diffusing particles.
The ensemble of absorbers we considered to have random walk
or polymer chain (self avoiding walk) statistics.
The flux of diffusing particles onto such kind of the absorber and its
higher moments generate a multifractal measure.\cite{Hentschel83}
The MF properties of this measure we described by
the spectral function formalism.\cite{Halsey86}

We performed our calculations in the frames of a field theoretical
approach, relating our study to the study of scaling properties of
composite field operators and defining their spectrum.
Namely we show that the ``copolymer star'' \cite{FerHol96e,FerHol96c}
- star operators $\prod_a^{f_1}\prod_b^{f_2}\psi_a\phi_b$ in a field
theory with interactions $u_\psi$ and $u_{\psi\phi} $ generate a
spectrum of scaling dimensions which transforms to a MF spectrum with
the appropriate convexity property. To calculate this spectrum we have
used the massless renormalization group scheme and massive
renormalization group approach at fixed dimension.

We give the explicit expressions for different kinds of exponents
describing our problem (formulas (\ref{3.23})-(\ref{3.26})),
(\ref{alpharwe}), (\ref{alpharwt}),
(\ref{alphasawe}), (\ref{alphasawt})
as well as
for the spectral function (\ref{frwe}), (\ref{fsawe}),
(\ref{frwt}), (\ref{fsawt}) in terms of
power series in
$\varepsilon=4-d$ and in pseudo-$\varepsilon$ expansion. All
calculations were performed  in the third order of perturbation
theory. In particular, in the second order in $\varepsilon$ we recover
previously obtained results.\cite{Cates87}

Special attention was payed to the fact that the series are
asymptotic and have zero radius of convergence. We have
used Pad\'e approximants to obtain analytic continuation of the series
under consideration for non-zero value if the expansion parameter.
In addition we applied resummation techniques well approved in field
theoretic calculations in order to obtain reliable information for the spectral
function $f(\alpha)$, H\"older exponent $\alpha$ and exponents
$\lambda$ governing scaling behavior of averaged density moments of
diffusing particles (see figs. \ref{fig2}a, \ref{fig2}b and table
\ref{tab2}). While standard in field theoretical
studies of critical phenomena, the resummation technique as to our
knowledge was not applied in the theory of multifractals. We hope that
our attempt will attract attention for this possibility in the context
of other problems arising in the theory of multifractal measures as well
as that the presumed accuracy of our results might evoke comparable efforts
by numerical simulation.
Further studies devoted to the set of spectra associated with
diffusion near the core of absorbing stars with higher numbers of arms.
We hope to gain more
insight on the unsymmetric behavior of the spectral function and find
the envelope of the family of spectra.

\section*{Acknowledgements}
We thank Lothar Sch\"afer, Universit\"at Essen, for valuable discussions. 
C.v.F. acknowledges support from the Minerva Gesellschaft.

%%%%%%%%%%%%%%%% TABLES and FIGURES
\begin{table} [H]
\caption {
\label{tab1}
Fixed points for the interactions of a system of polymers of two
species.
}
\begin{centering}
 \begin{tabular}{lllllllll}
 \\
 \hline
&$G_0$ & $U_0$ & $U_0^{\prime}$ & $S_0$ & $G$ & $U$ & $U^{\prime}$ &
$S$ \\
$g_{11}$ & 0 & $g^*$ & 0 &  $g^*$ & 0 &  $g^*$ & 0 &  $g^*$ \\
$g_{22}$ & 0 & 0 & $g^*$ & $g^*$ & 0 &  0 & $g^*$ &  $g^*$ \\
$g_{12}$ & 0 & 0 & 0 & 0 & $g^*_G$ & $g^*_U$ & $g^*_U$ &  $g^*$ \\
\hline
 \end{tabular}
\end{centering}
\end{table}

\begin{table} [H]
\caption {
\label{tab2}
Exponents $\lambda_{RW}(n)$ and $\lambda_{SAW}(n)$ obtained in
$\varepsilon$ and in pseudo-$\varepsilon$ expansion techniques.
}
\begin{centering}
\begin{tabular}{llllll}
\\
\\
\hline
\\
&$n$
& $\lambda_{RW}(\varepsilon)$ & $\lambda_{RW}(\tau)$
& $\lambda_{SAW}(\varepsilon)$ & $\lambda_{SAW}(\tau)$  \\
&1 &  0.99 &  0.99 & 0.71 & 0.71 \\
&2 &  1.77 &  1.81 & 1.31 & 1.33 \\
&3 &  2.45 &  2.53 & 1.86 & 1.92 \\
&4 &  3.01 &  3.17 & 2.34 & 2.44 \\
&5 &  3.51 &  3.75 & 2.78 & 2.94 \\
&6 &  3.95 &  4.28 & 3.19 & 3.41 \\
\\
\hline
\end{tabular}
\end{centering}
\end{table}

\end{document}